\documentclass[letterpaper]{article} 
\usepackage{aaai2026}  
\usepackage{times}  
\usepackage{helvet}  
\usepackage{courier}  
\usepackage[hyphens]{url}  
\usepackage{graphicx} 
\usepackage{natbib}  
\usepackage{caption} 
\usepackage{amsmath}   
\usepackage{amsfonts}  
\usepackage{amssymb}   
\usepackage{algorithm}
\usepackage{algorithmic}
\usepackage{array}
\usepackage{booktabs}
\usepackage{amsmath}
\usepackage{multirow}
\usepackage{newfloat}
\usepackage{listings}
\DeclareCaptionStyle{ruled}{labelfont=normalfont,labelsep=colon,strut=off} 
\floatstyle{ruled}
\newfloat{listing}{tb}{lst}{}
\floatname{listing}{Listing}
\title{Fine-Grained Image Quality Assessment for Perceptual Image Restoration}
\author{
    Xiangfei Sheng\equalcontrib\textsuperscript{\rm 1},
    Xiaofeng Pan\equalcontrib\textsuperscript{\rm 1},
    Zhichao Yang\textsuperscript{\rm 1},
    Pengfei Chen\textsuperscript{\rm 1},
    Leida Li\textsuperscript{\rm 1,2}\thanks{Corresponding author}
}
\affiliations{
    \textsuperscript{\rm 1}School of Artificial Intelligence, Xidian University \\
    \textsuperscript{\rm 2}State Key Lab. of Electromechanical Integrated Manufacturing of High-Performance Electronic Equipments, \\ Xidian University

    xiangfeisheng@gmail.com,
\{panxf@stu., yangzhichao@stu., chenpengfei@, ldli@\}xidian.edu.cn
}

\usepackage{bibentry}

\begin{document}

\maketitle

\begin{abstract}

Recent years have witnessed remarkable achievements in perceptual image restoration (IR), creating an urgent demand for accurate image quality assessment (IQA), which is essential for both performance comparison and algorithm optimization. Unfortunately, the existing IQA metrics exhibit inherent weakness for IR task, particularly when distinguishing fine-grained quality differences among restored images. To address this dilemma, we contribute the first-of-its-kind fine-grained image quality assessment dataset for image restoration, termed \textbf{FGRestore}, comprising 18,408 restored images across six common IR tasks. Beyond conventional scalar quality scores, FGRestore was also annotated with 30,886 fine-grained pairwise preferences. Based on FGRestore, a comprehensive benchmark was conducted on the existing IQA metrics, which reveal significant inconsistencies between score-based IQA evaluations and the fine-grained restoration quality. Motivated by these findings, we further propose \textbf{FGResQ}, a new IQA model specifically designed for image restoration, which features both coarse-grained score regression and fine-grained quality ranking. Extensive experiments and comparisons demonstrate that FGResQ significantly outperforms state-of-the-art IQA metrics.

\end{abstract}

\begin{links}
\link{Homepage}{https://sxfly99.github.io/FGResQ-Home/}
\end{links}

\section{Introduction}

Perceptual image restoration (IR) stands as a cornerstone in low-level computer vision, aiming to recover high-quality images from their degraded observations while maintaining perceptual fidelity~\cite{potlapalli2023promptir, luo2023controlling}. Recent years have witnessed remarkable breakthroughs in this field, largely driven by the evolution of generative models. These advances have enabled IR algorithms to achieve unprecedented visual quality, creating an urgent demand for accurate image quality assessment (IQA) methods that can reliably evaluate and compare restored images. Such assessment capabilities are essential not only for performance benchmarking across different IR algorithms but also for guiding algorithmic optimization.

\begin{figure}[t]
  \centering
  \includegraphics[width=0.78\linewidth]{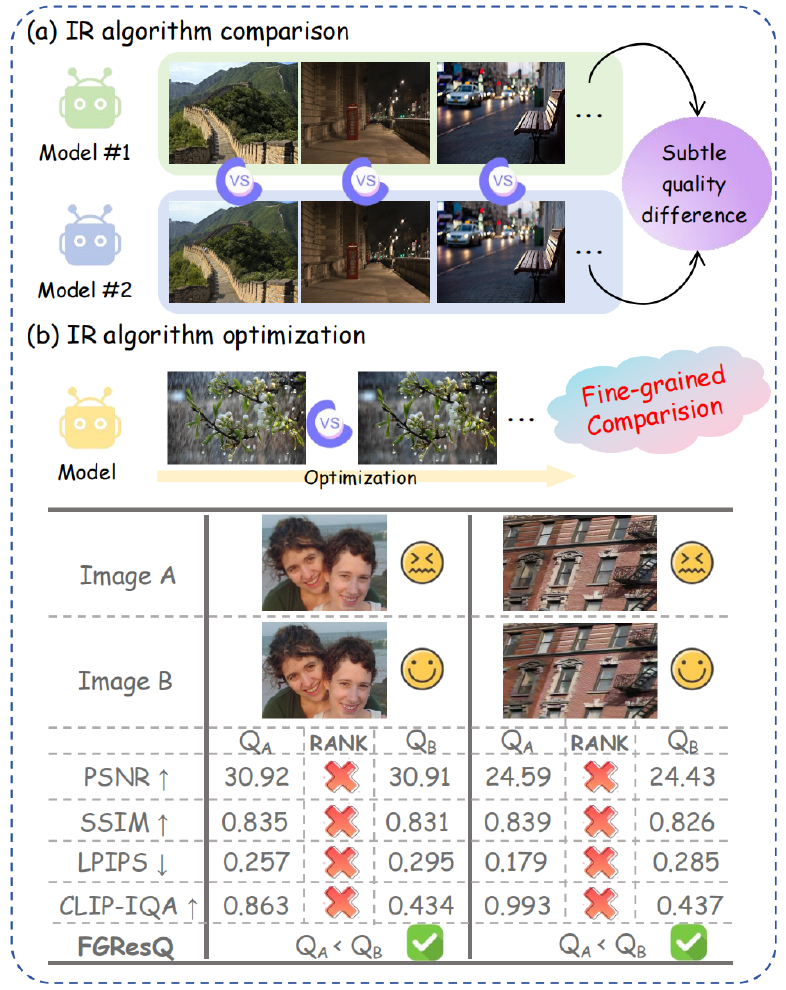}
\caption{Illustration of the \textbf{fine-grained challenge in IQA for image restoration}. Both IR algorithm comparison and optimization processes require distinguishing subtle quality differences between restored images. Existing IQA metrics fail to provide correct rankings for fine-grained image pairs. (Best viewed zoomed in.)}
  \label{fig:1}
\vspace{-3ex}
\end{figure}

\begin{figure*}[!t]
  \centering
  \includegraphics[width=0.95\linewidth]{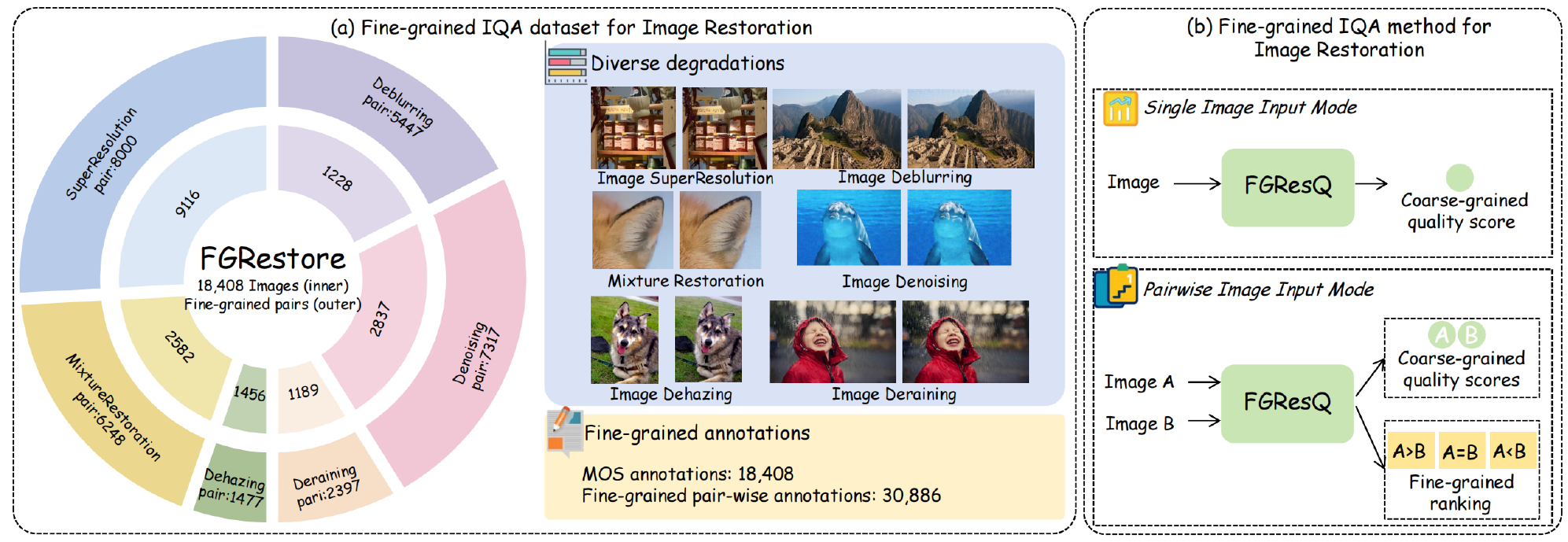}
\caption{Overview of our method. (a) \textbf{FGRestore} provides comprehensive fine-grained quality annotations for multiple IR tasks. (b) \textbf{FGResQ} enables both coarse-grained quality scoring and fine-grained quality ranking capabilities.}
  \label{fig:2}
\vspace{-3ex}
\end{figure*}

Currently, IR algorithm evaluation still predominantly relies on traditional reference-based metrics such as PSNR and SSIM~\cite{SSIM}, which assess restoration quality by measuring the similarity between the restored images and ground-truth images, even though their inconsistency with human perceptual judgment has been increasingly observed by the research community~\cite{PIPAL}. Recent efforts have introduced no-reference IQA methods~\cite{xu2024towards, zhou2025unires}, including CLIP-IQA~\cite{CLIP-IQA} and other learning-based approaches~\cite{MUSIQ}, which prove particularly valuable in real-world restoration scenarios where ground truth images are unavailable. However, a critical oversight persists: these methods have been primarily designed for image quality assessment, without specific consideration for the fine-grained evaluation requirements inherent in IR tasks.

A critical yet underexplored challenge in IR evaluation lies in its inherently \textbf{fine-grained nature}. As illustrated in Figure~\ref{fig:1}, IR algorithm comparison and optimization typically involve distinguishing between images with subtle quality differences—scenarios where even state-of-the-art IQA methods struggle to provide reliable assessments. This observation leads us to fundamentally rethink existing evaluation paradigms for IR tasks and raises a crucial question: 

\textit{Can existing score-based IQA methods objectively capture fine-grained quality differences in restored images?}

To investigate this question, we first conducted comprehensive computational analyses on established IQA datasets for image restoration~\cite{PIPAL,SISRSet}. Our findings consistently reveal a significant performance gap: while existing IQA methods achieve reasonable results in coarse-grained quality assessment, they consistently underperform in fine-grained scenarios, with significant inconsistencies between predicted scores and human perceptual judgments. This observation highlights the inadequacy of current evaluation protocols for capturing the nuanced quality differences which are crucial in IR applications. Moreover, when quality differences between images become subtle, both human and IQA models encounter significant challenges in providing reliable absolute quality scores. In contrast, pairwise comparison\cite{10168279,yang2024semantics} presents a more reliable alternative for fine-grained assessment, as humans demonstrate superior consistency in relative quality judgments compared to absolute scoring.

Motivated by the above insights, we introduce \textbf{FGRestore}, the first fine-grained IQA dataset specifically for image restoration tasks, as shown in Figure~\ref{fig:2}(a). Our dataset comprises 18,408 restored images spanning six common IR tasks: image super-resolution, deblurring, denoising, dehazing, deraining, and mixture restoration. Beyond conventional Mean Opinion Score (MOS) annotations, FGRestore incorporates 30,886 fine-grained preference annotations for restored image pairs with subtle quality differences. Based on FGRestore, we further propose \textbf{FGResQ}, a unified fine-grained IQA model tailored for IR evaluation (Figure~\ref{fig:2}(b)). Our approach employs degradation-aware feature learning that incorporates human-level degradation knowledge from Vision-Language Models (VLMs) into the proposed IQA framework, enabling both accurate quality scoring and fine-grained quality ranking across diverse IR tasks.

Our main contributions are threefold: 
\begin{itemize}
    \item We construct \textbf{FGRestore}, the first fine-grained IQA dataset specifically designed for IR tasks, containing 18,408 restored images across six common IR categories with both MOS and 30,886 fine-grained pairwise preference annotations.
    
    \item We propose \textbf{FGResQ}, a unified fine-grained IQA model that incorporates degradation-aware feature learning and demonstrates state-of-the-art performance in both coarse-grained score regression and fine-grained ranking.
    
    \item Through comprehensive computational analysis on existing IR datasets and extensive experiments, we reveal fundamental limitations of current score-based IQA methods in capturing fine-grained quality differences, providing valuable insights for future IR evaluation research.
\end{itemize}

\section{Related Work}
\subsection{Image Restoration Quality Assessment Dataset}

To evaluate IR algorithms, numerous datasets have been proposed. These include datasets dedicated to single restoration types, such as the MDD13~\cite{MDD13} dataset for deblurring, the IVC-Dehazing~\cite{IVC-Dehazing} dataset for dehazing, the IVIPC-DQA~\cite{IVIPC_DQA} dataset for deraining, and the QADS~\cite{QADS} and 
SRIQA-Bench~\cite{chen2025toward} datasets for super-resolution. Additionally, there are datasets that encompass distortions from various algorithms, such as PIPAL~\cite{PIPAL}, which includes distortion types like denoising, super-resolution, and super-resolution with post-denoising, KADID-10k~\cite{KADID-10k}, which contains various image distortions including denoising degradations and DiffIQA\cite{chen2025toward}, which accounts for degradation artifacts induced by diffusion-based image enhancement. A brief comparison of FGRestore and previous IR evaluation datasets is illustrated in Table 1.

\begin{table}[t]
\centering
\scriptsize
\renewcommand{\arraystretch}{0.85}
\setlength{\tabcolsep}{4pt} 
\begin{tabular}{p{1.5cm}>
{\centering\arraybackslash}p{0.4cm}>
{\centering\arraybackslash}p{0.3cm}>
{\centering\arraybackslash}p{0.3cm}>{\centering\arraybackslash}p{1.5cm}>{\centering\arraybackslash}p{2.5cm}}

\toprule
\multirow{2}{*}{\textbf{Dataset}} & \multirow{2}{*}{\textbf{Year}} & \multirow{2}{*}{\textbf{Task}} & \multirow{2}{*}{\textbf{Alg.}} & \textbf{Annotation} & \textbf{Number} \\
 &  &  &  & \textbf{Type} & \textbf{(Img./Pairs/Ann.)} \\
\midrule
MDD13 & 2013 & 1 & 5 & MOS (PC) & 1,200/0/13,592 \\
IVC-Dehazing & 2015 & 1 & 8 & MOS (ACR) & 200/0/4,800 \\
QADS & 2019 & 1 & 21 & MOS (PC) & 1,260/0/126,000 \\
SISRSet & 2019 & 1 & 8 & MOS (PC) & 260/0/32,000 \\
IVIPC-DQA & 2019 & 1 & 6 & MOS (5-level) & 2,136/0/27,192 \\
KADID-10k & 2019 & 1 & 1 & MOS(DCR) & 10,125/0/303,750 \\
exBeDDE & 2020 & 1 & 10 & MOS (PC) & 1,670/0/18,380 \\
PIPAL & 2020 & 3 & 40 & MOS (Elo) & 29,000/0/1.13M \\
RealSRQ & 2022 & 1 & 10 & MOS (PC) & 1,620/0/65,400 \\
DiffIQA & 2025 & 1 & 1 & RANK & 177,319/177,319/537,624 \\
SRIQA-Bench & 2025 & 1 & 10 & RANK & 1100/5500/55000 \\
\midrule
FGRestore & 2025 & 6 & 108 & MOS+RANK & 18,408/30,886/45,318 \\
\bottomrule
\end{tabular}
\caption{Comparison with the previous datasets. MOS: Mean Opinion Score; PC: pairwise comparison; ACR: Absolute Category Rating; DCR: Degradation Category Rating; 5-level: 5-point quality scale. Number show Images/Pairs/Annotations.}
\label{tab:source_datasets}
\vspace{-3ex}
\end{table}

\subsection{Image Quality Assessment}
Existing image restoration (IR) algorithms typically employ full-reference image quality assessment (FR-IQA) metrics for performance evaluation, such as the conventional PSNR and SSIM~\cite{SSIM}, as well as deep learning-based metrics like LPIPS~\cite{LPIPS} and DISTS~\cite{DISTS}. To overcome the flawed "perfect reference" assumption in traditional FR-IQA, which is challenged by imaging system limitations and superior generative methods, A-FINE~\cite{chen2025toward} proposes a generalized model that adaptively assesses both the fidelity and naturalness of test images.

No-reference IQA (NR-IQA) methods have been proposed to assess image quality without reference images. Early NR-IQA methods, such as NIQE\cite{NIQE}, primarily rely on hand-crafted Natural Scene Statistics (NSS) features. With the advancement of deep learning, many works have turned to neural networks for NR-IQA. For instance, MetaIQA\cite{MetaIQA}leverages meta-learning to enhance model generalization. Additionally, methods such as CLIP-IQA\cite{CLIP-IQA}leverage text prompts to enhance perceptual capabilities\cite{sheng2023aesclip,yang2025language,li2025cgiaa}. Given the powerful learning capabilities of large multi-modality models (LMMs)\cite{sheng2025instructcrop,sheng2025agiqa}, emerging work, such as Q-Align\cite{Q-align}, has begun to apply these models to quality assessment. Most recently, methods represented by Co-Instruct\cite{Co-Instruct}have extended LMMs to compare multiple images. Furthermore, VisualQuality-R1\cite{wu2025visualquality} introduces a novel methodology for IQA that leverages LMMs within a reinforcement learning-to-rank framework.

Although these datasets and methods have contributed significantly to the IQA community, most methods are score-based and cannot effectively handle fine-grained perceptual differences brought by image restoration algorithms. Recent works have explored fine-grained quality assessment in specific domains. Zhang et al.~\cite{zhang2019fine,zhang2022fine} introduced fine-grained assessment for compressed images, focusing on distinguishing compression artifacts within similar quality levels. However, these prior works primarily target image compression, whereas our work specifically addresses the unique challenges of fine-grained quality assessment in image restoration tasks, where the evaluation focus shifts from compression artifacts to perceptual restoration quality differences.

\begin{table}[t]
\centering
\small
\resizebox{\columnwidth}{!}{%
\begin{tabular}{llcccccc}
\toprule
\textbf{Type} & \textbf{Method} & \textbf{[0.0,0.2)} & \textbf{[0.2,0.4)} & \textbf{[0.4,0.6)} & \textbf{[0.6,0.8)} & \textbf{[0.8,1.0]} & \textbf{Overall} \\
& & \textbf{SRCC} & \textbf{SRCC} & \textbf{SRCC} & \textbf{SRCC} & \textbf{SRCC} & \textbf{SRCC} \\
\midrule
\multirow{4}{*}{\textbf{FR}} & PSNR & 0.323 & 0.082 & 0.209 & 0.161 & 0.072 & 0.422 \\
& SSIM & 0.293 & 0.108 & 0.258 & 0.254 & 0.049 & 0.530 \\
& LPIPS & -0.034 & 0.077 & 0.325 & 0.287 & 0.124 & 0.612 \\
& DISTS & 0.168 & 0.159 & 0.310 & 0.242 & 0.165 & 0.585 \\
\midrule
\multirow{10}{*}{\textbf{NR}} & NIQE & -0.126 & -0.002 & 0.107 & 0.001 & 0.080 & 0.153 \\
& IL-NIQE & -0.235 & -0.098 & 0.126 & 0.128 & 0.054 & 0.289 \\
& BRISQUE & -0.142 & 0.025 & 0.125 & 0.035 & 0.131 & 0.185 \\
& DB-CNN & -0.157 & 0.321 & 0.353 & 0.330 & -0.016 & 0.636 \\
& HyperIQA & 0.100 & 0.274 & 0.314 & 0.292 & 0.032 & 0.584 \\
& MetaIQA & 0.037 & 0.160 & 0.204 & 0.174 & -0.101 & 0.423 \\
& LIQE & -0.232 & 0.053 & 0.175 & 0.299 & 0.107 & 0.479 \\
& CLIP-IQA & -0.152 & 0.211 & 0.238 & 0.293 & 0.071 & 0.530 \\
& Q-Align & 0.230 & 0.301 & 0.337 & 0.213 & 0.178 & 0.418 \\
& DeQA-Score & 0.568 & 0.676 & 0.623 & 0.516 & 0.350 & 0.747 \\
\bottomrule
\end{tabular}%
}
\caption{Performance comparison across different MOS ranges on PIPAL dataset.}
\vspace{-3ex}
\end{table}

\begin{figure*}[t]
  \centering
  
  \includegraphics[width=1.0\linewidth]{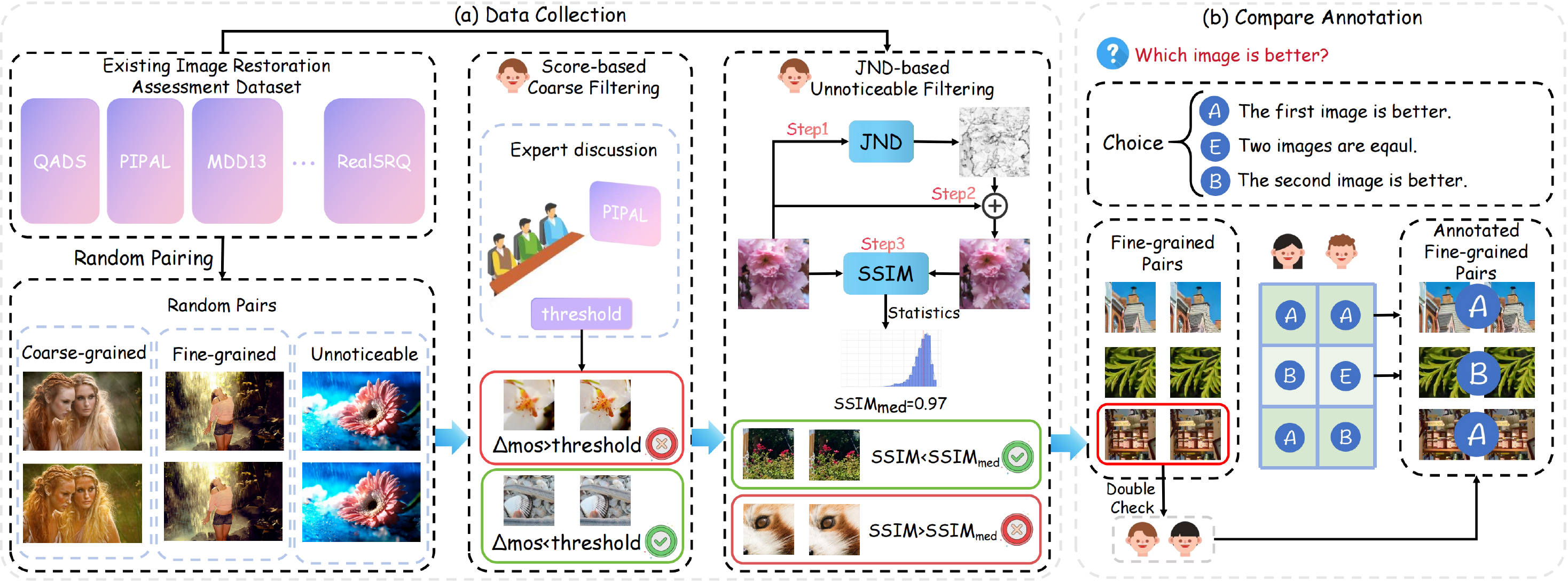}
\caption{Overview of the FGRestore dataset construction.}
  \label{fig:3}
  \vspace{-3ex}
\end{figure*}

\section{Preliminary Validation Analysis}

In image restoration tasks, both algorithm comparison and optimization frequently involve evaluating images with subtle quality differences. Algorithm comparison requires distinguishing between restoration results with marginal quality differences, while parameter optimization involves incremental quality changes that demand sensitive assessment methods to identify optimal configurations. To investigate whether existing IQA methods can objectively capture fine-grained quality differences in IR task, we conduct a comprehensive computational analysis on established IQA datasets. Specifically, we evaluate state-of-the-art IQA methods to assess their fine-grained discrimination capabilities. Our evaluation encompasses both full-reference (FR) and no-reference (NR) methods. FR methods include PSNR, SSIM~\cite{SSIM}, LPIPS~\cite{LPIPS}, and DISTS~\cite{DISTS}. NR methods include traditional approaches such as NIQE~\cite{NIQE}, IL-NIQE~\cite{IL-NIQE}, and BRISQUE~\cite{BRISQUE}, as well as deep learning-based methods including DB-CNN~\cite{DBCNN}, HyperIQA~\cite{HyperIQA}, MetaIQA~\cite{MetaIQA}, LIQE~\cite{LIQE}, CLIP-IQA~\cite{CLIP-IQA}, Q-Align~\cite{Q-align}, and DeQA-Score~\cite{DeQA}. 

Specifically, we partition the quality score range into several intervals to examine how IQA methods perform when evaluating images with similar quality levels, which is particularly relevant for fine-grained assessment scenarios in image restoration. 
Table~2 presents the performance comparison across different MOS ranges on the PIPAL dataset~\cite{PIPAL}. While most IQA methods achieve reasonable overall SRCC values on the whole dataset, their performance dramatically deteriorates when evaluated within narrow quality ranges. We also conducted similar analyses on other IQA datasets, which consistently demonstrate the same pattern. More results are provided in the \textit{supplementary material}. These findings provide strong evidence for the inadequacy of current score-based IQA approaches in fine-grained scenarios, motivating the development of our fine-grained evaluation framework for image restoration.

\section{FGRestore Dataset}
\label{sec:dataset}

Motivated by the findings revealed in our preliminary analysis, we construct FGRestore, the first comprehensive fine-grained image quality assessment dataset for perceptual image restoration evaluation. 
The dataset construction pipeline is illustrated in Figure~\ref{fig:3}, which involves collecting images from multiple datasets, filtering image pairs through score-based and JND-based criteria to retain fine-grained pairs, and conducting systematic subjective studies for pairwise preference annotation.

\subsection{Image and Pair Collection}

We collect images from multiple restoration-specific datasets to cover different degradation and visual appearances: 1) deraining images from IVIPC-DQA~\cite{IVIPC_DQA}; 2) deblurring images from MDD13~\cite{MDD13}; 3) dehazing images from IVC-Dehazing~\cite{IVC-Dehazing} and exBeDDE~\cite{exBeDDE}; 4) denoising images from PIPAL~\cite{PIPAL} and KADID-10k~\cite{KADID-10k}; 5) super-resolution images from QADS~\cite{QADS}, RealSRQ~\cite{RealSRQ}, SISRSet~\cite{SISRSet}, and PIPAL~\cite{PIPAL}; 6) mixture restoration images from PIPAL~\cite{PIPAL}.  After image collection, we generate image pairs by randomly pairing all images with identical content and restoration tasks, resulting in a total of 1,275,297 image pairs. 

\subsection{Data Filtration}

The randomly generated image pairs can be categorized into three distinct types based on perceptual quality differences: (1) \textit{Coarse-grained pairs} with highly noticeable quality differences, (2) \textit{Unnoticeable pairs} with negligible quality differences, and (3) \textit{Fine-grained pairs} with subtle but perceptible quality differences. We employ a two-step filtration process to remove coarse-grained and unnoticeable pairs.

\subsubsection{Score-based Coarse-grained Pairs Filtering.}

To eliminate coarse-grained pairs, we establish score difference thresholds for each source dataset through expert discussion. The coarse-grained filtering criterion is defined as:
{\footnotesize
\begin{equation}
\mathcal{F}_{\text{coarse}}(p_i) = \begin{cases}
1, & \text{if } |s_i^A - s_i^B| \leq \tau_d \\
0, & \text{otherwise}
\end{cases}
\label{eq:coarse_filtering},
\end{equation}
}
\noindent where $p_i = (I_i^A, I_i^B)$ represents the $i$-th image pair, $s_i^A$ and $s_i^B$ are the quality scores of images $I_i^A$ and $I_i^B$ respectively, and $\tau_d$ is the dataset-specific threshold. Pairs with $\mathcal{F}_{\text{coarse}}(p_i) = 1$ are retained for further processing.

\subsubsection{JND-based Unnoticeable Pairs Filtering.}

To remove unnoticeable pairs, we propose a JND-based filtering approach. The JND represents the minimum perceptual threshold for noticeable quality differences, making it ideal for identifying imperceptible quality pairs.
Our JND-based filtering process operates as follows: (1) randomly select 20,000 images from the collected datasets, (2) compute the JND map for each image using established JND estimation model~\cite{wu2017enhanced}, and (3) generate JND noise images by adding these JND maps as noise to the original images:
{\footnotesize
\begin{equation}
I_{\text{JND}} = I + \text{JND}(I)
\label{eq:jnd_overlay},
\end{equation}
}
\noindent where $I$ is the original image and $\text{JND}(I)$ is the corresponding JND noise map.
We calculate SSIM~\cite{SSIM} between each $I$ and $I_\text{JND}$, and use the median SSIM value $\text{SSIM}_{\text{med}}$ as our filtering threshold.
The unnoticeable pair filtering criterion is then defined as:
{\footnotesize
\begin{equation}
\mathcal{F}_{\text{unnotice}}(p_j) = \begin{cases}
1, & \text{if } \text{SSIM}(I_j^A, I_j^B) \leq \text{SSIM}_{\text{med}} \\
0, & \text{otherwise}
\end{cases}
\label{eq:unnoticeab_filtering},
\end{equation}}
\noindent where $p_j = (I_j^A, I_j^B)$ represents an image pair that passed the coarse-grained filtering. Pairs with $\mathcal{F}_{\text{unnotice}}(p_j) = 1$ are retained as fine-grained pairs.

After applying both filtering steps, we retain 30,886 fine-grained image pairs and their corresponding 18,408 images, forming the core of our FGRestore dataset. Due to space limitations, high-resolution fine-grained restoration examples are provided in the \textit{supplementary material}.

\begin{figure}[t]
\centering
\includegraphics[width=0.95\columnwidth]{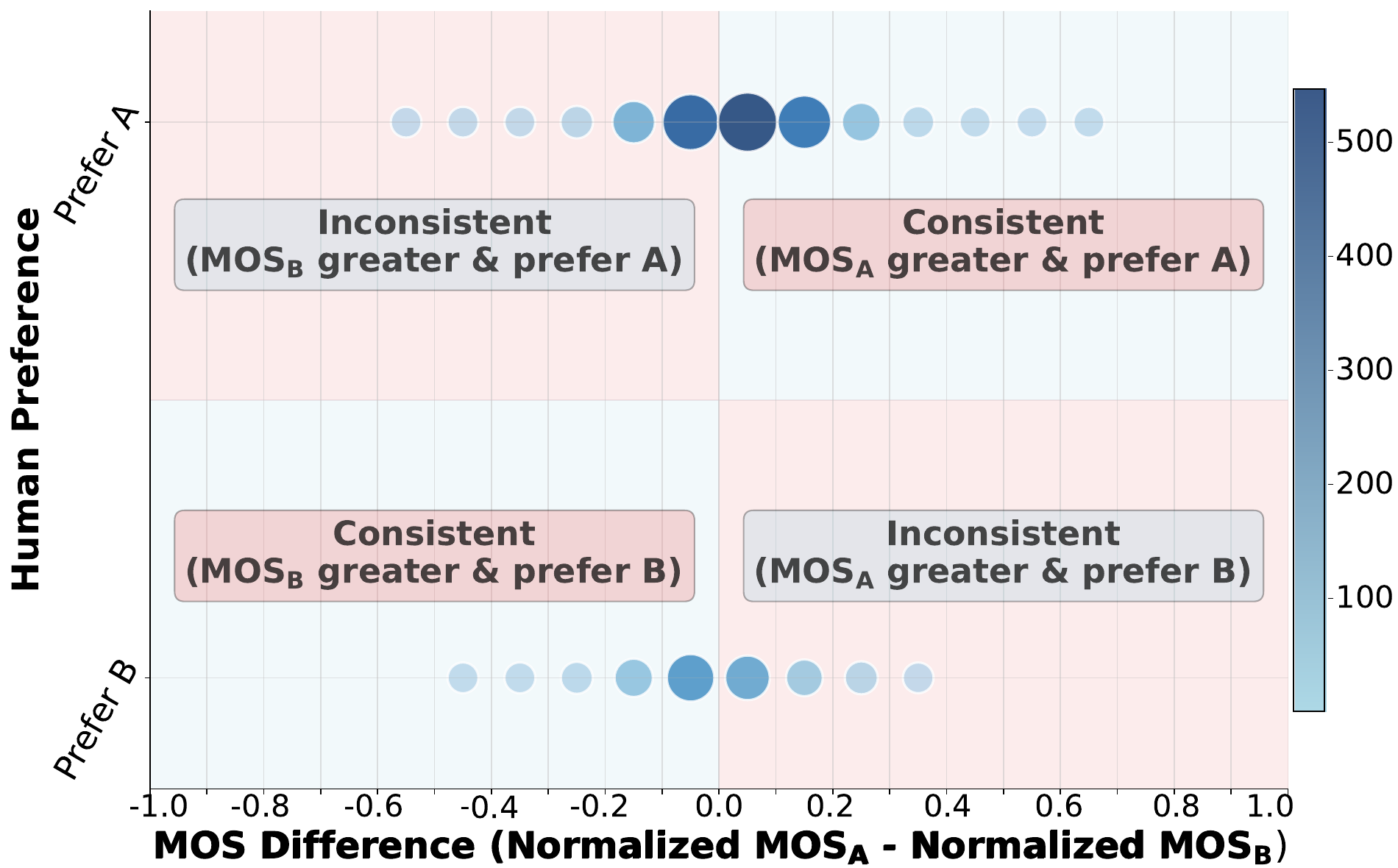}
\caption{Consistency analysis between MOS scores and human preference rankings. Point sizes represent frequency of image pairs. Red regions indicate inconsistent cases where MOS scores and human preferences disagree. }
\label{fig:consistency_analysis}
\vspace{-3ex}
\end{figure}

\subsection{In-lab Subjective Study}
FGRestore preserves the original score annotations from source datasets while introducing comprehensive fine-grained pairwise ranking annotations. For datasets originally annotated through pairwise ranking methodologies, fine-grained rankings can be directly derived from existing score relationships without additional human annotation. We design a resource-efficient two-round annotation protocol that incorporates quality control mechanisms to ensure reliable annotations. The annotation process is structured as follows:

\textbf{Annotation Protocol}. (1) Image pairs are divided into two groups, with each group assigned to a team of five trained annotators. For each image pair, annotators select the image with superior perceptual quality. Recognizing that some pairs may contain imperceptible differences despite filtering, annotators can also indicate equal quality when no clear preference exists. (2) Due to the subjective characteristics of fine-grained quality assessment, disagreements inevitably arise between annotators when quality differences are particularly subtle. Inconsistent annotations are resolved through expert review.
This systematic annotation approach yields 45,318 pairwise preference annotations, providing a robust foundation for fine-grained quality assessment model training and evaluation.
\subsection{Dataset Analysis}
To validate the necessity of fine-grained pairwise ranking, we analyze the consistency between MOS-based scoring and human preferences. Figure~\ref{fig:consistency_analysis} compares normalized MOS differences ($x$-axis) with human preferences ($y$-axis), revealing significant inconsistencies. Points in the red regions represent cases where MOS scores and preferences disagree. Critically, image pairs with smaller MOS differences exhibit larger scatter, indicating that inconsistencies worsen as quality differences become subtler. This supports the necessity of fine-grained pairwise ranking, as MOS scoring becomes unreliable for discriminating subtle quality differences.

\begin{figure*}[t]
\centering
\includegraphics[width=1.0\textwidth]{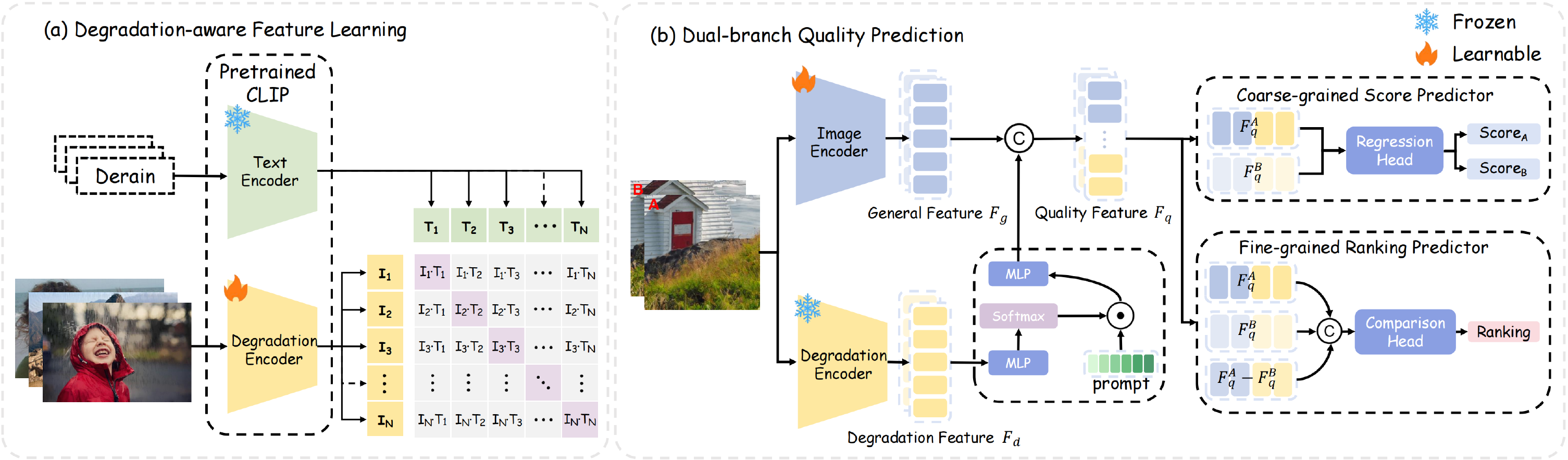}
\caption{Overview of the proposed FGResQ framework. }
\label{fig:framework}
\vspace{-3ex}
\end{figure*}

\section{FGResQ}
\label{sec:method}
Based on FGRestore, we propose FGResQ, a new fine-grained image quality assessment model for perceptual image restoration evaluation. The overall framework pipeline is illustrated in Figure~\ref{fig:framework}, which consists of two main components: (a) Degradation-aware Feature Learning that incorporates restoration task knowledge to enable unified evaluation across multiple IR tasks, and (b) Dual-branch Quality Prediction that simultaneously handles both coarse-grained score regression and fine-grained pairwise ranking.

\subsection{Degradation-aware Feature Learning}
\label{sec:degradation_aware}

To achieve unified quality assessment across diverse IR tasks, we propose a degradation-aware feature learning approach that enables our model to perceive and incorporate task-specific degradation characteristics. This design allows FGResQ to handle multiple IR scenarios within a unified framework, eliminating the need for task-specific model training. Specifically, we leverage a pretrained CLIP model to establish semantic alignment between visual content and degradation type. Specifically, we freeze the text encoder while fine-tuning a degradation encoder to learn degradation-aware representations. The degradation encoder maps images to a feature space that distinguishes different restoration scenarios. The learning objective employs bidirectional contrastive alignment to ensure robust degradation type perception:
{\footnotesize
\begin{equation}
\mathcal{L}_{cont} = \frac{1}{2}\left[\mathcal{L}_{CE}(\mathbf{F}_I \mathbf{F}_T^T, \mathbf{y}) + \mathcal{L}_{CE}(\mathbf{F}_T \mathbf{F}_I^T, \mathbf{y})\right],
\end{equation}}
\noindent where $\mathbf{F}_I$ and $\mathbf{F}_T$ represent image and text features respectively, $\mathcal{L}_{CE}$ denotes the cross-entropy loss, and $\mathbf{y} = [0, 1, 2, \ldots, N-1]$ represents the ground-truth matching labels for $N$ samples in a batch.


\subsection{Dual-branch Quality Prediction}
\label{sec:dual_branch}

Given degradation-aware features from the previous module, we first enhance them with learnable prompt embeddings to obtain quality-aware representations, then employ specialized prediction heads for different assessment tasks. Given an input image $I$, we extract general features $\mathbf{F}_g \in \mathbb{R}^d$ using the image encoder and degradation features $\mathbf{F}_d \in \mathbb{R}^d$ from the frozen degradation encoder. To effectively utilize degradation information, we employ learnable prompt embeddings $\mathbf{p}\in\mathbb{R}^{d}$ to transform degradation features into quality-aware features:
{
\begin{equation}
\begin{aligned}
\mathbf{F}_p &= \text{MLP}_1(\text{softmax} (\text{MLP}_2(\mathbf{F}_d))\cdot \mathbf{p}) \\
\mathbf{F}_q &= \mathbf{F}_g \oplus \mathbf{F}_p \oplus (\mathbf{F}_g+\mathbf{F}_p) \\
\end{aligned},
\end{equation}}
where $\text{MLP}_1$ and $\text{MLP}_2$ are multi-layer perceptrons, $\oplus$ denotes concatenation, and $\mathbf{F}_q$ represents the final quality-aware features.

FGResQ employs two specialized prediction heads: \textit{Regression Head} processes the quality features $\mathbf{F}_q$ to predict absolute quality scores $y_{pred}$ for individual images, enabling coarse-grained quality assessment compatible with traditional evaluation protocols. \textit{Comparison Head} handles fine-grained quality ranking by processing image pairs $(I_A, I_B)$ to obtain their respective quality features $(\mathbf{F}_{q}^A, \mathbf{F}_{q}^B)$ and predicting the ranking probability $p_{AB}$, directly addressing the fine-grained discrimination challenges identified in our preliminary analysis.
\subsubsection{Training Objectives}
We design a comprehensive training strategy through carefully designed loss functions:

\noindent\textbf{Scene-aware Fidelity Loss.} Since FGRestore preserves original score annotations from different source datasets with varying quality scales, we employ a scene-aware fidelity loss that focuses on ranking relationships within each dataset (scene) rather than absolute score values:
{\footnotesize
\begin{align}
    \mathcal{L}_{fid}^{(s)} = \frac{1}{\binom{N_s}{2}} \sum_{i<j} \Big( & 1 - \sqrt{p_{ij} g_{ij} + \epsilon} \nonumber \\
    & - \sqrt{(1-p_{ij})(1-g_{ij}) + \epsilon} \Big),
    \label{eq:your_label_here}
\end{align}}

where $p_{ij} = \sigma(y_{pred,i} - y_{pred,j})$, $g_{ij} = \frac{1}{2}(\text{sign}(y_{gt,i} - y_{gt,j}) + 1)$, and $N_s$ is the number of samples in scene $s$.

The overall scene-aware loss combines fidelity losses from different scenes with inverse sample count weighting:
{\footnotesize
\begin{equation}
\mathcal{L}_{scene} = \sum_{s \in \mathcal{S}} w_s \mathcal{L}_{fid}^{(s)},
\end{equation}}
where $w_s$ is the inverse sample count weight for scene $s$.
\textbf{Pairwise Ranking Loss.} For fine-grained pairwise preferences, we employ binary cross-entropy loss:
{\footnotesize
\begin{equation}
\mathcal{L}_{rank} = -\frac{1}{M} \sum_{i=1}^{M} [r_i \log(p_i) + (1-r_i) \log(1-p_i)],
\end{equation}}
where $M$ is the number of image pairs, $r_i$ is the ground-truth ranking label, and $p_i$ is the predicted ranking probability.
The overall training objective combines both components:
{\footnotesize
\begin{equation}
\mathcal{L}_{total} = \lambda_1 \mathcal{L}_{scene} + \lambda_2 \mathcal{L}_{rank},
\end{equation}}
where $\lambda_1$ and $\lambda_2$ are balancing hyperparameters.

\begin{table*}[t]
\centering
\renewcommand{\arraystretch}{0.75}
\fontsize{8}{9}\selectfont
\begin{tabular}{lp{2.0cm}p{1.455cm}*{9}{p{0.65cm}}}
\toprule
\multirow{2}{*}{\textbf{Type}} & \multirow{2}{*}{\textbf{Method}} & \multirow{2}{*}{\textbf{Pub.}} & \multicolumn{3}{c}{\textbf{Deblurring}} & \multicolumn{3}{c}{\textbf{Denoising}} & \multicolumn{3}{c}{\textbf{Deraining}} \\
\cmidrule(lr){4-6} \cmidrule(lr){7-9} \cmidrule(lr){10-12}
& & & \textbf{SRCC} & \textbf{PLCC} & \textbf{ACC} & \textbf{SRCC} & \textbf{PLCC} & \textbf{ACC} & \textbf{SRCC} & \textbf{PLCC} & \textbf{ACC} \\
\midrule
\multirow{4}{*}{\textbf{FR}} & PSNR & - & 0.187 & 0.167 & 0.634 & 0.487 & 0.482 & 0.775 & - & - & - \\
& SSIM & TIP'04 & 0.441 & 0.348 & 0.695 & 0.642 & 0.652 & \textbf{0.789} & - & - & - \\
& LPIPS &  CVPR'18 & 0.776 & 0.700 & 0.755 & 0.673 & 0.680 & 0.765 & - & - & - \\
& DISTS &  TPAMI'20 & 0.907 & \underline{0.901} & 0.845 & 0.679 & 0.672 & 0.739 & - & - & - \\ 
& A-FINE & CVPR'25 & \underline{0.907} & 0.796 & \underline{0.865} & 0.624 & 0.620 & 0.747 & - & - & - \\ 
\midrule
\multirow{8}{*}{\textbf{NR}} & NIQE & SPL'12 & 0.382 & 0.410 & 0.529 & 0.240 & 0.151 & 0.564 & 0.030 & 0.057 & 0.694 \\
& BRISQUE & TIP'12 &0.354 & 0.354 & 0.555 & 0.254 & 0.114 & 0.569 & 0.112 & 0.102 & 0.603 \\
& DB-CNN & TCSVT'22 &0.788 & 0.786 & 0.688 & 0.478 & 0.431 & 0.611 & 0.243 & 0.259 & 0.437 \\
& HyperIQA & CVPR'25 & 0.871 & 0.887 & 0.402 & 0.605 & 0.625 & 0.675 & 0.264 & 0.294 & 0.484 \\
& CLIP-IQA &  AAAI'23 &0.867 & 0.785 & 0.765 & 0.474 & 0.440 & 0.625 & 0.241 & 0.221 & 0.349 \\
& Q-Align & ICML'24 & 0.767 & 0.804 & 0.795 & 0.676 & 0.687 & 0.731 & 0.433 & 0.421 & 0.455 \\
& DeQA-Score & CVPR'25 & 0.815 & 0.843 & 0.819 & \underline{0.754} & \underline{0.771} & \underline{0.778} & \textbf{0.507} & \textbf{0.576} & 0.426 \\
& Compare2Score & NeurIPS'24 & 0.769 & 0.813 & 0.757 & 0.661 & 0.679 & 0.687 & 0.074 & 0.108 & \textbf{0.790} \\
\midrule
&FGResQ & - & \textbf{0.926} & \textbf{0.910} & \textbf{0.873} & \textbf{0.759} & \textbf{0.777} & 0.760 & \underline{0.496} & \underline{0.518} & \underline{0.778} \\
\midrule
\end{tabular}

\vspace{-0.15cm}
\centering
\begin{tabular}{l*{12}{p{0.65cm}}}
\midrule
\multirow{2}{*}{\textbf{Method}} & \multicolumn{3}{c}{\textbf{Dehazing}} & \multicolumn{3}{c}{\textbf{MixtureRestoration}} & \multicolumn{3}{c}{\textbf{SuperResolution}} & \multicolumn{3}{c}{\textbf{Average}} \\
\cmidrule(lr){2-4} \cmidrule(lr){5-7} \cmidrule(lr){8-10} \cmidrule(lr){11-13}
& \textbf{SRCC} & \textbf{PLCC} & \textbf{ACC} & \textbf{SRCC} & \textbf{PLCC} & \textbf{ACC} & \textbf{SRCC} & \textbf{PLCC} & \textbf{ACC} & \textbf{SRCC} & \textbf{PLCC} & \textbf{ACC} \\
\midrule
PSNR & - & - & - & 0.280 & 0.248 & 0.643 & 0.296 & 0.303 & 0.624 & 0.313 & 0.300 & 0.669 \\
SSIM & - & - & - & 0.421 & 0.379 & 0.684 & 0.351 & 0.361 & 0.641 & 0.464 & 0.435 & 0.702 \\
LPIPS & - & - & - & 0.466 & 0.475 & 0.658 & 0.448 & 0.460 & 0.666 & 0.591 & 0.579 & 0.711 \\
DISTS & - & - & - & 0.495 & 0.464 & 0.644 & 0.482 & 0.488 & 0.658 & 0.640 & 0.631 & 0.721 \\
A-FINE & - & - & - & 0.536 & 0.528 & 0.645 & 0.421 & 0.434 & 0.688 & 0.622 & 0.594 & \underline{0.736} \\ 
\midrule
NIQE & 0.036 & 0.063 & 0.446 & 0.041 & 0.027 & 0.541 & 0.176 & 0.072 & 0.503 & 0.151 & 0.130 & 0.546 \\
BRISQUE & 0.132 & 0.088 & 0.436 & 0.201 & 0.064 & 0.575 & 0.142 & 0.040 & 0.514 & 0.199 & 0.127 & 0.542 \\
DB-CNN & 0.643 & 0.645 & 0.524 & 0.415 & 0.460 & 0.614 & 0.459 & 0.454 & 0.643 & 0.504 & 0.506 & 0.586 \\
HyperIQA & 0.643 & 0.674 & 0.409 & 0.523 & 0.535 & 0.499 & 0.437 & 0.433 & 0.538 & 0.557 & 0.574 & 0.501 \\
CLIP-IQA & 0.547 & 0.499 & 0.546 & 0.302 & 0.300 & 0.580 & 0.244 & 0.186 & 0.571 & 0.446 & 0.405 & 0.573 \\
Q-Align & 0.715 & 0.765 & 0.584 & 0.569 & 0.571 & 0.658 & 0.376 & 0.366 & 0.662 & 0.589 & 0.603 & 0.648 \\
DeQA-Score & \underline{0.762} & \underline{0.803} & \underline{0.644} & \underline{0.669} & \underline{0.679} & \underline{0.697} & \textbf{0.561} & \textbf{0.573} & \underline{0.718} & \underline{0.678} & \underline{0.707} & 0.680 \\
Compare2Score & 0.334 & 0.381 & 0.436 & 0.494 & 0.519 & 0.635 & 0.317 & 0.315 & 0.607 & 0.441 & 0.469 & 0.652 \\
\midrule
\textbf{FGResQ} & \textbf{0.821} & \textbf{0.854} & \textbf{0.669} & \textbf{0.698} & \textbf{0.706} & \textbf{0.713} & \underline{0.521} & \underline{0.536} & \textbf{0.721} & \textbf{0.703} &  \textbf{0.717}  & \textbf{0.752}  \\
\bottomrule
\end{tabular}
\caption{Performance comparison on FGRestore dataset across different IR tasks. "-" indicates no reference images available.}
\label{tab:performance_comparison}
\vspace{-3ex}
\end{table*}

\section{Experiments}

\subsection{Evaluation Protocol} We compare the proposed FGResQ model against state-of-the-art FR and NR IQA methods. For fair comparison, all learning-based methods with public source codes are retrained using the same training-testing protocol with our scene-aware loss replacing their original regression loss, except for Compare2Score \cite{Compare2Score} which operates in a zero-shot manner without training. We evaluate performance using SRCC and PLCC for quality score prediction, and accuracy (ACC) for fine-grained pairwise ranking. We split the dataset at an 8:2 ratio by image pairs. Individual image division following the corresponding pair.

\subsection{Implementation Details} We use CLIP ViT-B/16 as the backbone encoder for degradation-aware feature learning and dual-branch quality prediction. The model is trained on NVIDIA RTX4090 GPUs using the Adam optimizer with cosine annealing. The maximum learning rate is $5 \times 10^{-6}$ and the batch size is 64. To ensure training stability, we employ scene-aware sampling that groups samples from the same source dataset within each batch. $\lambda_1$ and $\lambda_2$ are set to 5 and 1, respectively. 

\subsection{Performance Evaluation}

\textbf{Overall Performance Analysis.} Table~\ref{tab:performance_comparison} presents comprehensive performance comparisons across all six IR tasks in our FGRestore dataset. FGResQ achieves state-of-the-art performance across most evaluation metrics, with our method demonstrating particularly significant advantages in fine-grained ranking accuracy (ACC), which serves as the most critical metric for fine-grained quality assessment. The substantial improvement in ACC demonstrates FGResQ's superior capability in distinguishing subtle quality differences. When compared to the strongest baseline methods, FGResQ shows consistent improvements across different evaluation paradigms. Among existing approaches, DeQA-Score exhibits the best coarse-grained regression performance, yet FGResQ surpasses it by significant margins with 2.5\% SRCC improvement and 7.2\% ACC improvement. More remarkably, our proposed method outperforms FR methods like DISTS despite these methods having access to additional reference images for quality assessment. This superior performance highlights the critical importance of explicitly modeling fine-grained quality relationships rather than relying solely on reference-based similarity measures, demonstrating that specialized fine-grained assessment capabilities are crucial for accurate IR evaluation.

\textbf{Fine-grained vs. Coarse-grained Assessment.} A critical observation is the performance gap between regression metrics and ranking accuracy across different methods. While some methods achieve reasonable regression performance, their fine-grained ranking capabilities remain limited.

\begin{figure}[!t]
\centering
\includegraphics[width=1.0\columnwidth]{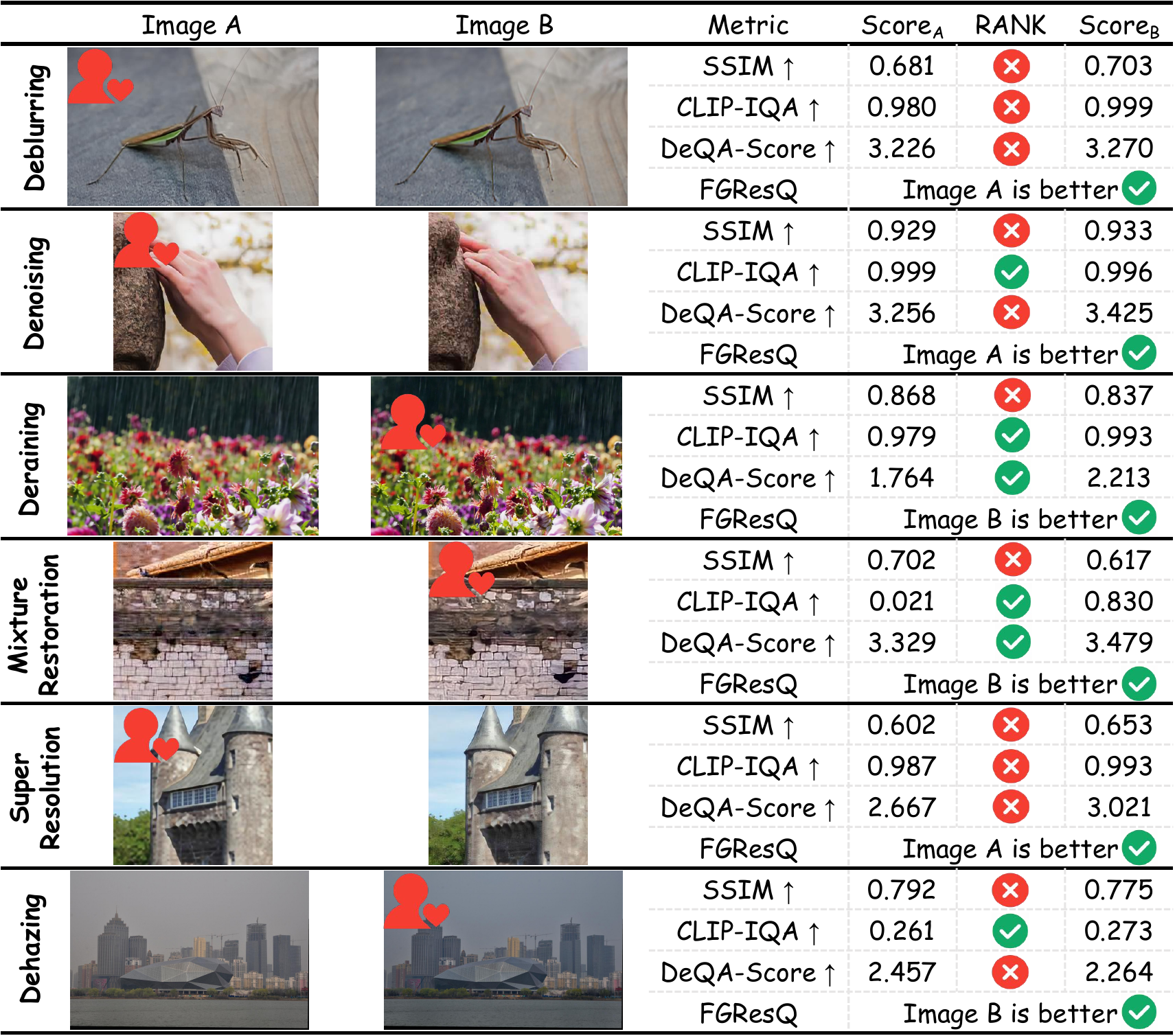}
\caption{Qualitative comparison. (Best viewed zoomed in.)}
\label{fig:example}
\vspace{-3ex}
\end{figure}

\textbf{Qualitative Analysis.} Figure~\ref{fig:example} presents qualitative comparisons on representative fine-grained image pairs across different restoration tasks. The results demonstrate systematic limitations of existing IQA methods in fine-grained scenarios. Traditional metrics like PSNR and SSIM often produce nearly identical scores, failing to provide meaningful quality discrimination. Advanced learning-based methods such as CLIP-IQA and DeQA-Score also struggle with subtle quality differences, frequently producing incorrect rankings. In contrast, FGResQ consistently identifies the superior image across tested cases, demonstrating the effectiveness of our pairwise comparison approach. Additional qualitative examples are provided in the \textit{supplementary material}.
\vspace{-3ex}
\section{Conclusion}

In this work, we have addressed a critical yet underexplored challenge in image restoration evaluation: the inadequacy of existing IQA methods for fine-grained quality assessment. To address this fundamental limitation, we introduced FGRestore, the first fine-grained IQA dataset specifically designed for IR evaluation, comprising 18,408 restored images across six restoration tasks with 30,886 fine-grained pairwise preference annotations. Based on this, our proposed FGResQ model achieves state-of-the-art with significant improvements in fine-grained ranking accuracy. This work establishes a new evaluation framework for image restoration and provides valuable insights for developing more perceptually-aligned quality assessment methods in the era of advanced generative restoration models.

\section{Acknowledgments}
This work is supported by National Natural Science Foundation of China under Grants 62471349, 62171340 and 62301378,  Fundamental Research Funds for the Central Universities under Grant QTZX25076, and partly supported by the China Postdoctoral Science Foundation under Grant 2024M762553.

\bibliography{aaai2026}

\end{document}